\documentclass[12pt,a4paper]{article}

\usepackage{graphicx}
\usepackage{setspace}	
\usepackage{amsmath}	
\usepackage{amssymb}	
\usepackage{titling}	
\usepackage{siunitx}	

\usepackage{color}	
\usepackage{hyperref}
\hypersetup{unicode=true,
	   colorlinks=true,
	   linkcolor=blue,
	   citecolor=red,
	   pdfborder={0 0 0}
}
\usepackage{dcolumn}
\newcolumntype{d}[1]{D{.}{.}{#1}}

\newcommand{\alfaAA}{\ensuremath\alpha_{\mathrm{AA}}}
\newcommand{\alfaAB}{\ensuremath\alpha_{\mathrm{AB}}}
\newcommand{\alfaBA}{\ensuremath\alpha_{\mathrm{BA}}}
\newcommand{\alfaBB}{\ensuremath\alpha_{\mathrm{BB}}}
\newcommand{\dd}{\ensuremath\mathrm{d}}
\newcommand{\Denom}{\ensuremath 1 + \alfaAA\RA + \alfaBB\RB + \gamma\RA\RB}
\newcommand{\eigE}{\ensuremath \mathcal{E}}
\newcommand{\EFermi}{\ensuremath E_\mathrm{F}}
\newcommand{\eV}{\ensuremath \mathrm{eV}}
\newcommand{\Goff}{\ensuremath G^\mathrm{off}}
\newcommand{\IA}{\ensuremath I_\mathrm{A}}
\newcommand{\IB}{\ensuremath I_\mathrm{B}}
\newcommand{\iterprima}{\textsc{iter1\;}}
\newcommand{\itersecunda}{\textsc{iter2\;}}
\newcommand{\itertertia}{\textsc{iter3\;}}
\newcommand{\iterquarta}{\textsc{iter4\;}}
\newcommand{\kO}{\ensuremath \mathrm{k}\Omega}
\newcommand{\Knum}{\ensuremath \mathcal{K}}
\newcommand{\nm}{\ensuremath \mathrm{nm}}
\newcommand{\NT}{\ensuremath N_\mathrm{T}}
\newcommand{\pA}{\ensuremath \mathrm{pA}}
\newcommand{\pts}{\hspace*{1pt}}
\newcommand{\RA}{\ensuremath R_\mathrm{A}}
\newcommand{\RAB}{\ensuremath R_\mathrm{A,B}}
\newcommand{\RB}{\ensuremath R_\mathrm{B}}
\newcommand{\tB}{\ensuremath t_\mathrm{B}}
\newcommand{\VA}{\ensuremath V_\mathrm{A}}
\newcommand{\VB}{\ensuremath V_\mathrm{B}}
\newcommand{\Vg}{\ensuremath V_\mathrm{g}}
\newcommand{\Vmax}{\ensuremath V_\mathrm{max}}

\begin{document}
\title{Twin lead ballistic conductor based on nanoribbon edge transport}
\author{%
	Martin Kon\^{o}pka* and Peter Die\v{s}ka
	\\ \\
	SLOVAK UNIVERSITY OF TECHNOLOGY in Bratislava,\\
	Faculty of Electrical Engineering and Information Technology,\\
	Institute of Nuclear and Physical Engineering,\\
	Department of Physics,\\
	Ilkovi\v{c}ova 3, 812~19 Bratislava, Slovak Republic
	\\
	\small{*E-mail: \href{mailto:martin.konopka@stuba.sk}{martin.konopka@stuba.sk}}}
\date{$31^{\,\textrm{st}}$ October 2017}
%
\maketitle
\begin{abstract}
If a device like a graphene nanoribbon (GNR) has all its four corners attached to electric current leads,
the device becomes a quantum junction through which two electrical circuits can interact.
We study such system theoretically for stationary currents.
The 4-point energy-dependent conductance matrix of the nanostructure and the classical resistors in the circuits are
parameters of the model.
The two bias voltages in the circuits are the control variables of the studied system while the electrochemical
potentials at the device's terminals are non-trivially dependent on the voltages.
For the special case of the linear-response regime analytical formulae for the operation of the coupled
quantum-classical device are derived and applied.
For higher bias voltages numerical solutions are obtained.
The effects of non-equilibrium Fermi levels are captured using a recursive algorithm in which self-consistency
between the electrochemical potentials and the currents is reached within few iterations.
The developed approach allows to study scenarios ranging from independent circuits to strongly coupled ones.
For the chosen model of the GNR with highly conductive zigzag edges we determine the regime in which the single device
carries two almost independent currents.
\end{abstract}
\begin{quote}
\end{quote}
%
%
\section{\label{sec:intro} Introduction}
%
%
Electronic transport through graphene nanoribbons has attracted attention in particular because their zigzag edges
exhibit magnetic properties and support specific modes, which may influence the charge
transport~\cite{Nakada1996,Yazyev2008,Niu2009}.
Edges of larger-scale graphene ribbons have been proposed to serve similarly or analogously to
optical fibres or waveguides~\cite{Allen_2016}.
In experimental works studying electronic transport through a nanoribbon, the flake is typically contacted to
electrodes across its whole width, see for instance Ref.~\cite{Dai2008}.
However, if the edges are to be used for the transport, it may be appropriate to make contacts only to the corners of
such a ribbon.
This poses a question if a single graphene nanoribbon or an alternative planar atomistic structure could serve as two
wires carrying two independent electric currents.
Obviously, there would be some interaction between the currents flowing along the two parallel edges and they would not
be fully independent.
Still, if the energy ranges of the high density of states, resulting from the zigzag (ZZ) edge modes, are used for the
transport, the conductance along the ZZ direction may be significantly larger than the conductance along the
perpendicular (armchair, AC) direction.
In such case the two ZZ edges might be considered as two relatively independent spatially separated conductors.
More generally, such scenario would be the case if the $4 \times 4$ conductance matrix of a ribbon preferred one of the
directions over the perpendicular and diagonal directions.
On the other hand, a nanoribbon with significant conductance along several directions would allow to study the
intriguing regime in which the two classical circuits are coupled through the quantum ballistic device.
Regardless of the conductance characteristics, such setups would naturally require four electrodes contacted to the
corners of the nanoribbon.

Electronic transport through graphene flakes with electrodes contacted at their corners has been studied
computationally in Ref.~\cite{Konopka2015}, with focus on non-rectangular flakes such as trapezoids or triangles.
The work has addressed the question how the two-contact conductance qualitatively depends on the magnitude of the
angles in the two contacted corner areas and on the types of the edges forming the corner.
It was in addition found that the usual nearest-neighbor (NN) tight-binding (TB) model
was insufficient for proper description of the edge-induced transport properties
(provided that the electrodes were attached to the corners).

Four-terminal phase-coherent conductance has been considered by B\"{u}ttiker~\cite{Buttiker1986} in order to clarify
the occurrence of asymmetric magnetoresistances.
The author considered a sample contacted to two circuits, what is a setup similar to the one assumed in the present
work.
On the contrary, the model of B\"{u}ttiker does not include any classical resistors and is limited to the linear
regime.
We will return to it in the exposition of our theoretical description.
Four-terminal schemes have many times been studied both theoretically and experimentally in setups with two voltage
probes implying that two of the four terminals carry zero net currents.
Analysis of this specific model can be found as early as in the above cited work by B\"{u}ttiker.
A four-terminal electron waveguide coupler was proposed and theoretically analysed in Ref.~\cite{Wang1992},
assuming ballistic transport.
Based on the formulae of Ref.~\cite{Buttiker1986}, B\"{u}ttiker's resistance tensor of the about $500\,${\AA} long
junction has been calculated.
In Ref.~\cite{Hirayama1992} an experimental realisation of two parallel wires coupled by a ballistic window
was described.
Molecular logic gates with intramolecular circuits are another example of nanoscale multiterminal devices.
Considered schemes include both ballistic components as well as classical
resistors~\cite{Joachim_2002_PRB,Joachim_2002_Nanotech}.
Theoretical analysis and computational results on multiple-lead coherent conductors in case of finite applied voltages
were reported in Ref.~\cite{Lesovik_1993}.
A particularly studied case was a four-terminal conductor with two of the currents set to zero.
Voltage difference between the two open leads as a non-linear function of the current was calculated.

In this paper we study a coupled quantum-classical device consisting of a ballistic transport junction
and of two classical circuits with the four leads contacted at the junction.
As a concrete realisation of the junction a GNR is considered.
The two electric currents are driven by DC bias voltage sources in the circuits that include resistors.
For the linear regime, transparent analytical formulae displaying the interplay of the quantum and classical elements
are derived for both the electric currents as well as for the electrochemical or electrostatic
potentials in the device's leads.
A numerical scheme is described to tackle the non-linear regime of the composite system.
In order to include the effects of the non-equilibrium (NE) electrochemical potentials (EChPs)
on the current-voltage characteristics and other quantities,
a recursive method for the coupled device is developed.
Numerical demonstrations are performed for both a perfect GNR and for a perturbed one.
We show that under certain conditions the two electrical circuits of the scheme operate almost independently.
Hence such a GNR could serve as two relatively independent conductors, which might be found useful in experimental
setups and practical applications as it would allow to shrink the device size.
The regime in which the circuits are significantly coupled through the ballistic device
presents an interesting physical model and is cover by our numerical demonstrations as well.

In Sect.~\ref{sec:model} we define the system under study.
In Sect.~\ref{sec:theory} we derive main results of our work both for the linear and non-linear regimes.
We also provide additional description of the studied model, relevant for the NE calculations.
In Sect.~\ref{sec:applicate} we apply the theoretical modelling to study the setup defined in Sect.~\ref{sec:model}.
Additional details can be found in the Supplementary Material (SM)~\cite{SM}.
We conclude our results in Sect.~\ref{sec:concl}.
%
%
\section{\label{sec:model}The model and methods}
\begin{figure}[t]
\centerline{\includegraphics[width=0.40\columnwidth]{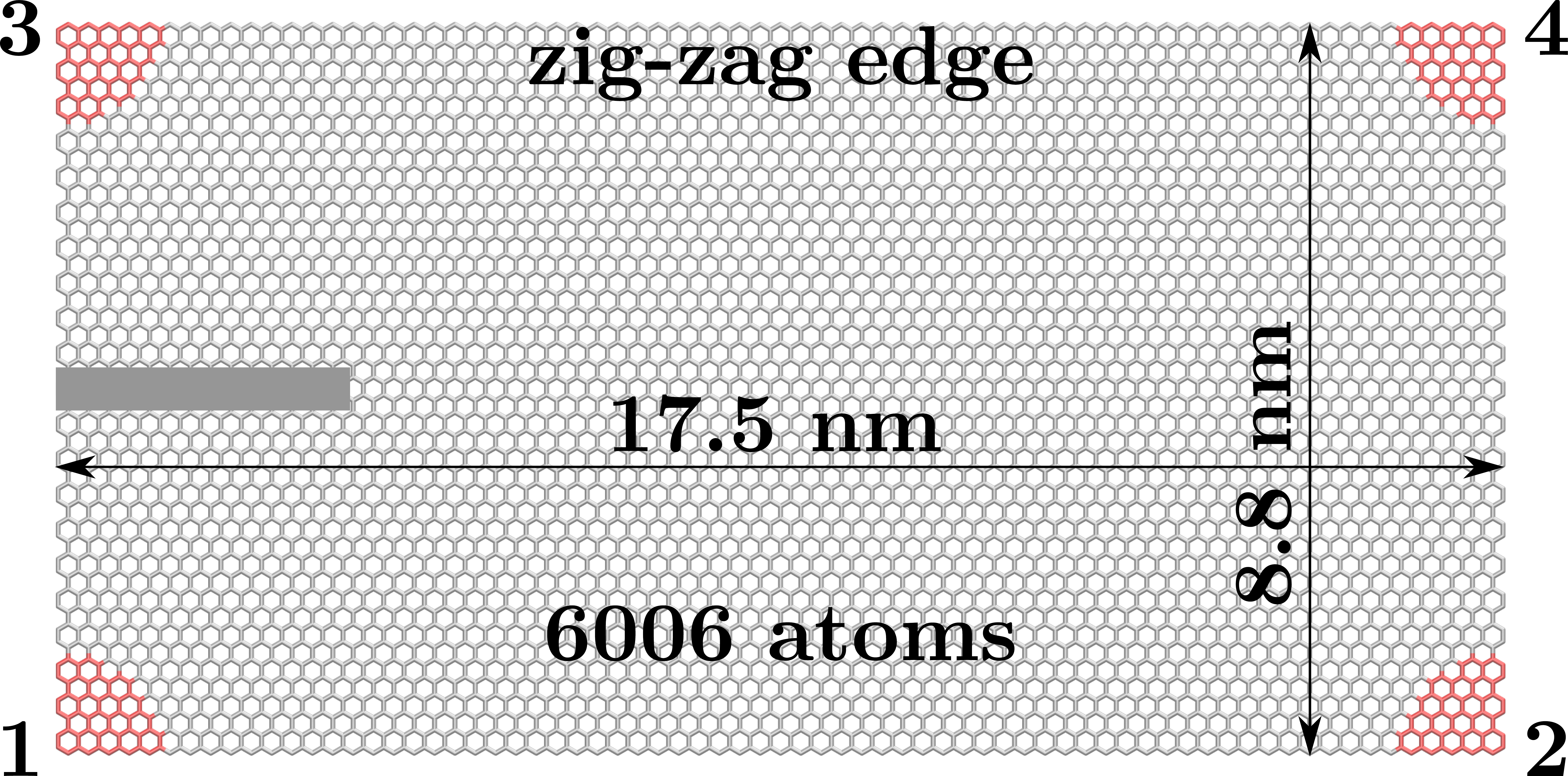}}
\caption{The graphene nanoribbon (GNR) with the four electrodes attached at its corners, which are labelled as
$1, \dots, 4$ for further reference.
Each electrode is modelled by a bunch of $54$ identical mono-atomically thin wires~\cite{Konopka2015} (not shown)
coupled to the red-marked carbon atoms.
The GNR's dimensions are $L = 71\,b$ (along the ZZ edges) and $W = 62\,a$ (along the AC edges),
with $a = 0.142\,\nm$ being the nearest-neighbor distance in graphene and $b = a \sqrt{3}$ the lattice parameter.
The GNR is composed of $6006$ atoms.
In an alternative adapted convention~\cite{Ruffieux2016} the GNR's dimensions are $(143,42)$.
The dark narrow strip on the left-hand side parallel to the ZZ edges marks those atoms ($60$ in total, grouped
in $15$ 4-atom segments) that are removed
from the perfect GNR to form the perturbed structure, which is also considered in the text.
A detail of a perturbed GNR is shown as the inset of Fig.~\ref{fig:VB_IA_IB_Equilib_Lin}(b).}
\label{fig:GNR}
\end{figure}

\begin{figure}[t]
\centerline{\includegraphics[width=0.60\columnwidth]{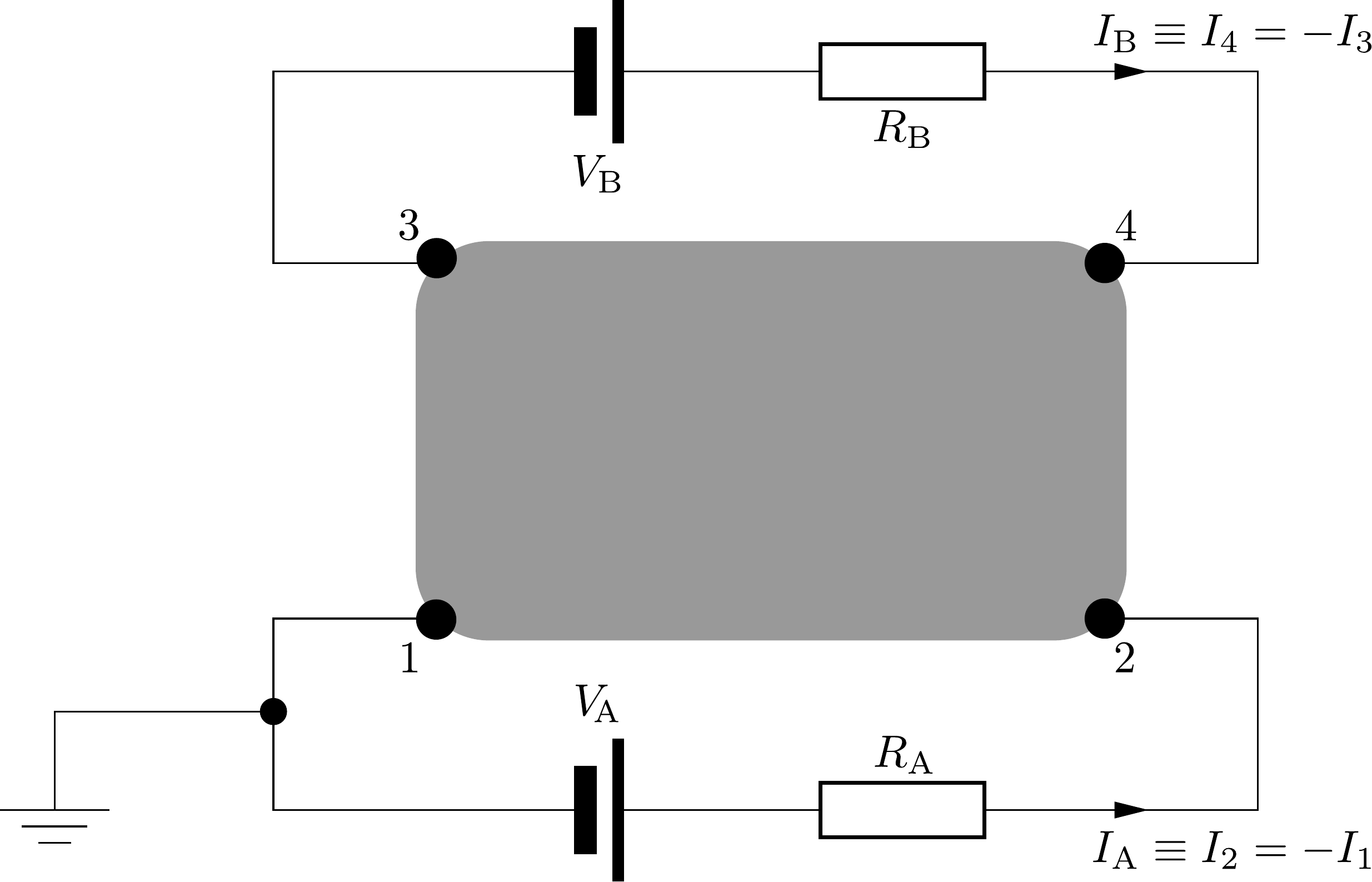}}
\caption{The electric scheme of the considered device.
Given control variables are the two bias voltages $\VA$ a $\VB$.
Additional parameters are the classical resistances $\RA$ and $\RB$ and the $4 \times 4$
conductance matrix $G$, which in general depends on the energy and on the EChPs of the leads.
Stationary regime is assumed for each element of the scheme.
The arrows on the leads mark the conventionally positive directions of the currents $\IA$ and $\IB$ in the circuits.}
\label{fig:scheme}
\end{figure}

%
\paragraph{The physical model.}
An atomistic model of the considered GNRs is shown in Fig.~\ref{fig:GNR}.
The red-marked carbon atoms in the corners form the contact areas where the four electrodes (not shown) are attached.
If this is accomplished, the whole system in a simple electrical setup can be described by the scheme drawn
in Fig.~\ref{fig:scheme}.
%
%
The resistances $\RA$ and $\RB$ of the two classical resistors are known parameters.
The control variables are the static bias voltages $\VA$ and $\VB$ on the two DC sources,
\emph{not} the EChPs in the leads.
The potentials as well as the two currents are unknown non-trivial functions of the bias voltages
and have to be determined as it is explained in detail below in the text.
The stationary regime is assumed for all elements of the scheme.
We consider a low-temperature regime in which inelastic scattering of the electrons in the GNR and in its contacts can
be neglected.
Hence the GNR with its contacts represents a quantum subsystem of the entire scheme.
The electronic transport properties of such a contacted device in the linear-response regime can be described
by a $4 \times 4$ conductance matrix $G$.
This matrix is computed using a microscopic model described below.
For sufficiently low bias voltages the linear regime with $\IA, \IB \propto \VA, \VB$ can be assumed
and the conductance matrix is then considered as energy-independent.
This will allow us to derive analytical formulae for currents $\IA(\VA,\VB)$ and $\IB(\VA,\VB)$ in terms of the fixed
system parameters $G$, $\RA$, and $\RB$.
%

We recall that in the scheme under study (Fig.~\ref{fig:scheme}) not only the currents but also the EChPs
$\mu_2$, $\mu_3$ and $\mu_4$ are unknown non-trivial functions, $\mu_\alpha(\VA,\VB)$,
which have to be determined and must strictly be distinguished from the bias voltages.
This requirement is dictated by the scheme itself in which the differences $\mu_2 - \mu_1$ and $\mu_4 - \mu_3$ are
\textit{a priori} unknown because the currents through the resistors are unknown.
The chemical potential at the terminal 1 is fixed by the ground-level condition.
One of the foci of the present work is the non-linear response regime with generally non-equilibrium
EChPs $\mu_\alpha$ in individual leads.
In the non-linear regime the response of the junction will be described by an energy-dependent conductance matrix
$G(\eigE)$ instead just a simple numerical matrix.
If, in addition, the NE values of $\mu_\alpha$'s are assumed, the conductance matrix will acquire the additional
dependences: $G = G(\eigE; \{\mu_{\alpha}\})$.
\paragraph{The microscopic description.}
We consider GNRs with the electrodes attached to their corners according to Fig.~\ref{fig:GNR}.
The model of the electrodes and the electronic structure description are similar to those used
in Ref.~\cite{Konopka2015}
with a notable exception of the more general model of the electrodes in the present work.
We use the independent-electron approximation and an extended tight-binding (TB) hamiltonian for the GNR
with interactions up to the 3rd NN included.
The hopping parameters $\tB = -2.97\,\eV$, $\tB' = -0.073\,\eV$ and $\tB'' = -0.33\,\eV$
are taken from Ref.~\cite{Reich2002}.
The orbital overlaps are neglected.
The extended interaction range within the GNR was found important to account for the contributions of the ZZ edge
state to the electronic transport if the electrodes are attached to corners of graphene flakes~\cite{Konopka2015}.
We do not consider the spin-related effects;
the focus of the present work is the description of a four-terminal nanoribbon device coupled to two classical branches
of the circuits.
A generalisation of our formalism including the spin effects at least on the level of the mean-field
approximation to Hubbard hamiltonian would be relatively straightforward although computationally much more
expensive~\cite{Yazyev2010}.
Quantitatively accurate modelling of GNR's electronic structure especially for narrow ribbons would also require
an inclusion of the electron-electron interactions into the model~\cite{Ruffieux2016}.

Each electrode in our model is formed by a bunch of mutually non-interacting monoatomically thin
wires~\cite{our_PRB_2014,Konopka2015} described within the $2$nd NN approximation with the couplings
$t_1 = \tB$ and $t_2 = \tB'$.
The particular model drawn in Fig.~\ref{fig:GNR} and used for our quantitative analyses employs $54$ wires per
electrode.
Using just a single wire would allow us to study very small currents only.
The introduction of the $2$nd NN couplings $t_2$ for the atoms of the wires is the only important modification of
the used microscopic theory in comparison to work~\cite{Konopka2015}.
The equilibrium on-site (atomic orbital) energies are denoted as $\epsilon$ and are assumed to be the same for the
entire system, including the electrodes.
The extended TB model of the monoatomically thin wires yields the one-electron dispersion relation
$\eigE(\Knum) = \epsilon + \sum_{n=1}^{N_\mathrm{far}} 2 t_n \cos(n\Knum)$,
with $N_\mathrm{far}$ being farthest nearest neighbor considered in the wires;
here $N_\mathrm{far} = 2$.
The Fermi energy of the wires is $\EFermi = \eigE(\pi/2) = \epsilon - 2 t_2$.
Introduction of the extended TB model of the electrodes thus causes the positive shift $-2 t_2$ of the Fermi energy
compared to the basic TB model.
We set $\epsilon = 2 t_2 = 2 \tB'$ what results in having $\EFermi = 0$.

The couplings between the wires and the GNR again extend up to the $2$nd NN
from the vertex atom of the GNR (the atom of the GNR to which a given wire is attached).
In other words, the vertex atom is coupled to two closest atoms of the semi-infinite wire.
Our model assumes the same coupling parameters $t_1$ and $t_2$ as defined above for the intra-wire interactions.

In the present work we pay attention to the non-linear response regime of the coupled quantum-classical device
including the transport with an account of NE Fermi levels.
Techniques to tackle these regimes, given the specifics of the studied scheme, are described
in Sects.~\ref{sssec:nonlin} and~\ref{sssec:NonEqTheory}.
In this way the specification of our model from Sect.~\ref{sec:model} will be supplemented to be usable also under the
NE conditions.
%
%
\section{\label{sec:theory}Theory}
%
%
\subsection{Quantum-mechanical calculations}
%
Standard Green's function formalism (described for instance in Ref.~\cite{Ryndyk})
is used to calculate eigenfunctions of the entire system.
Knowledge of the eigenfunctions allows us to compute the transmission and reflection coefficients for electrons incoming
from the wires.
With the exception of the extended TB model of the wires the calculations have been accomplished according to the route
described in Ref.~\cite{Konopka2015}.
The Green's function formalism requires knowledge of the semi-infinite wire's eigenstates which are provided in
the SM~\cite{SM}.
%
\subsection{Landauer-B\"{u}ttiker formula}
%
We will employ the well-known Landauer-B\"{u}ttiker (LB) type formula for stationary currents through the leads of
a multi-terminal device:
\begin{equation}
I_\alpha =
-\frac{2e}{h} \sum_{\substack{\beta=1 \\ \beta\ne\alpha}}^{\NT}
\int_{\mu_{\beta}}^{\mu_{\alpha}} \mathcal{T}_{\alpha\leftarrow\beta}(\eigE) \, \dd\eigE
\, , \ \ \ \alpha \in \{1, 2, \dots, \NT\}
\pts .
\label{eq:LB}
\end{equation}
$\NT$ is the number of the terminals (also the electrodes),
$\mathcal{T}_{\alpha\leftarrow\beta}(\eigE)$ is the transmission
coefficient from the terminal $\beta$ to the terminal $\alpha$
and $\mu_\gamma$ are the EChPs associated with the individual wires.
Formula~\eqref{eq:LB} assumes (i)~zero temperature, (ii)~the time-reversal symmetry (TRS)
and (iii)~the trivial description of the spin degrees of freedom (spin degeneracy).
The TRS implies that
\begin{equation}
\sum_{\beta (\ne\alpha)} \mathcal{T}_{\alpha\leftarrow\beta}(\eigE)
=
\sum_{\beta (\ne\alpha)} \mathcal{T}_{\beta\leftarrow\alpha}(\eigE)
\pts .
\label{eq:trs}
\end{equation}
We note that in our model each electrode (attached at its corresponding terminal) is assumed to be composed of a number
of the mono-atomically thin wires.
We assume that $\alpha$ and $\beta$ label the electrodes, not the wires.
Still, the TRS formula~\eqref{eq:trs} is equally exactly valid for these composite electrodes as it would be valid
if the indices $\alpha$ and $\beta$ labelled the individual wires attached on the central device.

The charge conservation for the central device implies Kirchhoff's first law:
\(\sum_{\alpha} I_\alpha = 0\).
The EChP $\mu_\alpha$ at the terminal $\alpha$ is conveniently expressed in terms of the
electrostatic potential $U_\alpha$ associated with the terminal (the electrode):
\begin{equation}
\label{eq:mu}
\mu_\alpha \equiv -e U_\alpha
\pts.
\end{equation}
We recall that
in our model (see its description in Sect.~\ref{sec:model} and in Fig.~\ref{fig:scheme}) the
EChPs themselves as well as the currents are non-trivial
unknown functions of the two bias voltages $\VA$ and $\VB$ as will be further explained below.
%
\subsection{Inclusion of the circuits with resistors into the 4-terminal device}
%
\subsubsection{\label{sssec:theproblem}Formulation of the problem}
Consider a device with 4 terminals, i.e. $\NT=4$ like that drawn in Fig.~\ref{fig:scheme}.
Each terminal has an electrode connected to it.
In the basic statement of the LB formula~\eqref{eq:LB} above the (stationary) currents  $I_\alpha$ obey
to Kirchhoff's first law and no other constraints were supposed.
Now we put in consequences the assumption that the four leads form the two circuits and there are also the two
resistors included so that the entire system can be schematically depicted as on Fig.~\ref{fig:scheme}.
The lower circuit (A) has one of its branches grounded.
For such 4-terminal device we have to require the following additional constraining condition,
\begin{equation}
I_3 + I_4 = 0
\pts ,
\label{eq:circ_34}
\end{equation}
which is more restrictive than just Kirchhoff's first law.
We assume the stationary regime for each element of the entire scheme, i.e. also for the sources.
Therefore the condition~\eqref{eq:circ_34} is necessary, otherwise an unphysical charge imbalance would be gradually
built on the voltage source $\VB$ and that would in addition contradict the assumption of the stationarity.
Kirchhoff's first law together with the Eq.~\eqref{eq:circ_34} then imply a similar
condition on the currents $I_1$ and $I_2$:
\begin{equation}
I_1 + I_2 = 0
\pts.
\label{eq:circ_12}
\end{equation}
The presence of the circuits requires that, in addition to the Eqs.~\eqref{eq:circ_34} and~\eqref{eq:circ_12},
Kirchhoff's second law has to be fulfilled:
\begin{subequations}
\label{eq:Kirchoff2}
\begin{align}
\label{eq:Kirchoff2_U2}
U_2 &= U_1 + \VA - \RA \IA \pts ,
\\
\label{eq:Kirchoff2_U4}
U_4 &= U_3 + \VB - \RB \IB \pts .
\end{align}
\end{subequations}
We recall the relation~\eqref{eq:mu}: $\mu_\alpha \equiv -e U_\alpha$.
According to the scheme in Fig.~\ref{fig:scheme}, we can choose
\begin{equation}
U_1 = 0
\label{eq:ground}
\end{equation}
without any significant loss of generality.
This auxiliary setting allows us to accomplish derivations using the potentials $U_2$, $U_3$ and $U_4$ instead
of the differences $U_2-U_1$, $U_3-U_1$ and $U_4-U_1$.
Hence the three electrostatic potentials $U_2$, $U_3$, $U_4$ and the four currents $I_\alpha$
are unknown quantities and have to be determined as functions of the two given bias voltages $\VA$ and $\VB$.
\subsubsection{\label{sssec:lin_nospin}The linearity approximation}
Under the assumption of the section's title, formula~\eqref{eq:LB} for the electric currents takes the
frequently accounted form
\begin{equation}
I_\alpha
\ = \
\sum_{\substack{\beta=1 \\ \beta\ne\alpha}}^{\NT} G_{\alpha\beta} \, (U_\alpha - U_\beta)
\pts .
\label{eq:I_G_U}
\end{equation}
We stress again that for the scheme under study (see Fig.~\ref{fig:scheme}) the electrostatic potentials $U_\gamma$
must not be reduced to or confused with the bias voltages $\VA$ and $\VB$.
See also the reasoning in Sect.~\ref{sec:model}.
The linear conductance matrix elements are $G_{\alpha\beta} \equiv (2 e^2/h) \, \mathcal{T}_{\alpha\leftarrow\beta}$.
The four equations~\eqref{eq:I_G_U} together with the constraints~\eqref{eq:circ_34}, \eqref{eq:circ_12} and
with the Eqs.~\eqref{eq:Kirchoff2} form an inhomogeneous set of eight linear equations with $7$ unknowns
$U_2$, $U_3$, $U_4$, $I_1$, $\dots$, $I_4$.
Known parameters are the two resistances $\RA$ and $\RB$, the matrix elements $G_{\alpha\beta}$ satisfying
the TRS of the form~\eqref{eq:trs} and the two bias voltages $\VA$ and $\VB$.
I.e. the two voltages, $\VA$ and $\VB$, are supposed to be the control variables in such an experiment,
\emph{not} the three electrostatic potentials $U_2$, $U_3$, and $U_4$.
There is a linear dependence among the 8 equations.
Once the set is properly reduced to remove the dependence, we obtain an inhomogeneous set with a unique solution.
The equations are solved in the SM~\cite{SM} and resulting linear relations $\IA = \IA(\VA,\VB)$ and
$\IB = \IA(\VA,\VB)$ are obtained:
\begin{subequations}
\label{eq:IA_IB_vs_VA_VB_lin}
\begin{align}
\IA &= \frac{\bigl(\alfaAA + \gamma\RB\bigr)\VA - \alfaAB\VB}{\Denom} \pts ,
\\[8pt]
\IB &= \frac{\bigl(\alfaBB + \gamma\RA\bigr)\VB - \alfaBA\VA}{\Denom} \pts ,
\end{align}
\end{subequations}
where $\alpha_{CC'}$ are conductances introduced by B\"{u}ttiker~\cite{Buttiker1986}
and depend solely on the conductance matrix elements.
For the problem in hand they take the form
$\alfaAA = S_1 - (G_{13}+G_{14}) (G_{31}+G_{41})/S$,
$\alfaAB = (G_{13} G_{24} - G_{14} G_{23})/S$,
$\alfaBA = (G_{31} G_{42} - G_{41} G_{32})/S$
and
$\alfaBB = S_3 - (G_{31}+G_{32}) (G_{13}+G_{23})/S$,
with
$S = G_{13}+G_{23}+G_{14}+G_{24}$
and
$S_\alpha = \sum_{\beta \ne \alpha} G_{\alpha\beta}$.
Alternative expressions are provided in the SM~\cite{SM}.
For convenience we have introduced the parameter
\begin{equation}
\label{eq:gamma}
\gamma = \alfaAA\alfaBB - \alfaBA\alfaAB
\pts .
\end{equation}
In the SM~\cite{SM} we find and present formulae for the all potentials differences $U_\alpha - U_\beta$.
Among the other differences, we find
\begin{align}
\nonumber
U_4 - U_2 = -\frac{1}{S D(G,R)} \,\bigl[
&\bigl(G_{31}+G_{41}\bigr)\VA + \bigl(G_{41} S_3 + G_{31} G_{43}\bigr)\RB\VA\bigr.
\\[-12pt]
\label{eq:U42_solo}
\\[-8pt]
\nonumber
\bigl.
- &\bigl(G_{13}+G_{23}\bigr)\VB - \bigl(G_{23} S_1 + G_{13} G_{21}\bigr)\RA\VB
\bigr],
\end{align}
as well as a more compact formula
$U_4 - U_3 = [(1+\alfaAA\RA)\VB + \alfaBA\RB\VA]/D$.
The shorthand notation
\begin{equation}
D(G,R) = D = \Denom
\label{eq:Denom}
\end{equation}
represents the commonly occurring denominator in the above expressions~\eqref{eq:IA_IB_vs_VA_VB_lin}
for the currents as well as in formula~\eqref{eq:U42_solo} for the differences of the potentials.
Expressions for the remaining differences $U_\alpha-U_\beta$ can be determined from the above stated formulae
using the symmetries of the considered scheme.

The Eqs.~\eqref{eq:IA_IB_vs_VA_VB_lin} and \eqref{eq:U42_solo} are our main analytical results for the linear regime,
together with similar equations for the remaining $U_\alpha - U_\beta$ differences~\cite{SM}.
The Eqs.~\eqref{eq:IA_IB_vs_VA_VB_lin} generalise B\"{u}ttiker's formulae~\cite{Buttiker1986}
and transparently show how the currents in the two circuits relate both to the classical
and to the quantum elements of the scheme.

The Eq.~\eqref{eq:U42_solo} describes how the potentials difference $U_4-U_2$ (and analogously $U_3-U_1$)
can be varied through a continuous range of values by simply tuning the classical resistors in the circuits.
For example,
we can obtain zero difference $U_4 - U_2$, i.e. also the equality $\mu_4 = \mu_2$ of the EChPs.
[See the relations~\eqref{eq:mu}.]
With one of the batteries reversed (for instance $\VB \to -\VB$), one can similarly try to set to zero at least one
of the potentials differences across the sample as it can be seen from the
results for the other $U_\alpha - U_\beta$ differences~\cite{SM}.
An illustration of the effect will be shown in Sect.~\ref{sssec:balance}.
Obviously, a particular condition like $U_4 = U_2$ is only achievable for proper combinations of the numerical values
of the quantities entering the relation.
\subsubsection{\label{sssec:nonlin}The 4-terminal device beyond the linear response}
In this section we provide a generalisation of the considerations in Sect.~\ref{sssec:lin_nospin} beyond the linear
regime.
It will allow us to more thoroughly describe nano-devices like junctions formed by graphene flakes that are often
found to operate in non-linear regimes.
The quantum-mechanical part of the system is parameterised by the matrix $\mathcal{T}$ standing in the
Eq.~\eqref{eq:LB}, which is now a set of (presumably known) functions of the energy.
The energy dependence is the substantial complication in comparison to the linear regime in which
the matrix elements were constants.
To make the formulae free of the numerical prefactors $2 e^2/h$, we equivalently use an energy-dependent conductance
matrix instead of $\mathcal{T}_{\alpha\leftarrow\beta}(\eigE)$.
To keep the dimensions correct, we introduce an auxiliary quantity
\begin{equation}
u \equiv \frac{\eigE}{e}
\, ,
\label{eq:u_E_e}
\end{equation}
where $e > 0$ is the unit of charge.
The conductance matrix elements are then written as
\begin{equation}
G_{\alpha\beta}(u) \equiv \frac{2e^2}{h} \, \mathcal{T}_{\alpha\leftarrow\beta}(e u)
\pts .
\label{eq:condGmat_E}
\end{equation}
The equations to be solved are~\eqref{eq:LB} with $\NT = 4$, again with the constraints~\eqref{eq:circ_34}
and~\eqref{eq:circ_12} and Kirchhoff's second law Eqs.~\eqref{eq:Kirchoff2}.
For convenience we collect and write down all these equations and conditions here,
making their specific to the 4-terminal system under study:
\begin{equation}
\begin{aligned}
-\sum_{\substack{\beta=1 \\ \beta\ne 1}}^4 \int_{\mu_{\beta}/e}^{\mu_1/e} G_{1\beta}(u) \, \dd u &= I_1 = -\IA \pts ,\\
-\sum_{\substack{\beta=1 \\ \beta\ne 2}}^4 \int_{\mu_{\beta}/e}^{\mu_2/e} G_{2\beta}(u) \, \dd u &= I_2 = +\IA \pts ,\\
-\sum_{\substack{\beta=1 \\ \beta\ne 3}}^4 \int_{\mu_{\beta}/e}^{\mu_3/e} G_{3\beta}(u) \, \dd u &= I_3 = -\IB \pts ,\\
-\sum_{\substack{\beta=1 \\ \beta\ne 4}}^4 \int_{\mu_{\beta}/e}^{\mu_4/e} G_{4\beta}(u) \, \dd u &= I_4 = +\IB \pts .
\end{aligned}
\label{eq:curr_4T_itg_set}
\end{equation}
Accompanying relations~\eqref{eq:Kirchoff2} and~\eqref{eq:ground} for the classical part of the system are now more
conveniently expressed in terms of the EChPs $\mu_\alpha \equiv -e U_\alpha$
rather than the electrostatic ones:
\begin{equation}
\mu_2 = \mu_1 - e \VA + e \RA \IA \, , \ \ \ \ \ \
\mu_4 = \mu_3 - e \VB + e \RB \IB \, , \ \ \ \ \ \
\mu_1 \equiv \EFermi = 0\, .
\label{eq:EChP_4T}
\end{equation}
The accomplish the task we have to solve the coupled set of the above equations~\eqref{eq:curr_4T_itg_set}
and~\eqref{eq:EChP_4T}, most of them being integral equations.
Given parameters and known functions are $\RA$, $\RB$, $\VA$, $\VB$ and $G_{\alpha\beta}(u)$.
Main unknown physical quantities to be determined are the currents $\IA$ and $\IB$.
Of secondary interest are the EChPs $\mu_\alpha$ which are also unknown with the exception of $\mu_1$.
Obviously, all this is a much more difficult task than it was in the case of its linear counterpart
in Sect.~\ref{sssec:lin_nospin}.
For the non-linear case we develop a numerical algorithm which takes $\mu_2$, $\mu_3$, $\mu_4$, $\IA$ and $\IB$
as unknown functions of the two independent (and presumably experimentally controllable) variables $\VA$ and $\VB$.
As the key part of the derivation,
we express the coupled set of the equations~\eqref{eq:curr_4T_itg_set} and~\eqref{eq:EChP_4T} also at varied values
$\VA + \delta\VA$ and $\VB + \delta\VB$ of the bias voltages.
Resulting dependent variables are $\mu_\alpha(\VA+\delta\VA, \VB+\delta\VB) = \mu_\alpha(\VA,\VB) + \delta\mu_\alpha$,
$\,\alpha \in \{2, 3, 4\}$, and analogously for the currents.
This differentiation leads to a set of algebraic equations of the form~\eqref{eq:I_G_U} and~\eqref{eq:Kirchoff2},
where instead of the variables stand their differentials~\cite{SM}.
As a result, $\IA(\VA,\VB)$, $\IB(\VA,\VB)$ and $\mu_{\alpha}(\VA,\VB)$ are obtained as some generally non-linear
functions for a given (\emph{fixed}) set of $G_{\alpha,\beta}(u)$ spectra, see the Eq.~\eqref{eq:condGmat_E}.
\subsubsection{\label{sssec:NonEqTheory}Non-equilibrium Fermi levels: recursive calculation of the non-linear response}
We now return to the Landauer-B\"{u}ttiker type formula~\eqref{eq:LB}.
Given bias voltages $(\VA, \VB)$ on the batteries induce a set of the \textit{a priori} unknown EChPs
$\mu_1, \dots, \mu_4$, in short $\mu$, which enter the formula~\eqref{eq:LB}, or, equivalently,
the Eqs.~\eqref{eq:curr_4T_itg_set}.
The transmission coefficients spectra $\mathcal{T}_{\alpha\leftarrow\beta}(\eigE)$
depend on the unknown EChPs:
$\mathcal{T}_{\alpha\leftarrow\beta} = \mathcal{T}_{\alpha\leftarrow\beta}(\eigE; \mu)$.
In terms of the conductance matrix~\eqref{eq:condGmat_E}, $G_{\alpha\beta} = G_{\alpha\beta}(u;\mu)$.
The dependence on the unknown $\mu$'s makes the analysis of the coupled device much more complicated even
in comparison to the non-linear regime treatment in Sect.~\ref{sssec:nonlin}.
The complications were not faced there thanks to the approximation consisting
in the use of the the single (fixed) set of spectra $G_{\alpha\beta}(u)$ assumed to be usable for any $\mu$,
hence for any $(\VA, \VB)$ within a studied domain.
Obviously, this approximative treatment in general would not yield correct results for the studied model.
It is an acceptable approximation for sufficiently low voltages;
numerical examples for different levels of approximation will be shown in Sect.~\ref{sssec:NonEqDemo}
and in the SM~\cite{SM}.
We note that non-equilibrium response in a weakly non-linear regime was discussed in Ref.~\cite{Christen1996}.
More general approaches to quantum transport including the non-equilibrium EChPs often employ
the non-equilibrium Green's function method, see for instance~\cite{HaugJauho}.
In the present study we use the independent-electron approximation as defined in Sect.~\ref{sec:model}.
In order to capture the effects of the non-equilibrium potentials $\mu$
(what we have not done until the present section)
we calculate the Green's functions~\cite{Ryndyk} using the non-equilibrium potential profile.
Given the specifics of the system under study (in particular the explicit presence of the circuits
and the non-trivial dependence of $\mu$'s on chosen bias voltages $\VA$ and $\VB$),
we proceed in the following steps:
\begin{enumerate}
\itemsep0pt
\item
In a quantum-mechanical (QM) calculation we obtain $\mathcal{T}_{\alpha\leftarrow\beta}(\eigE; \mu)$
with all $\mu$'s set to the equilibrium Fermi energy $\EFermi$.
We denote these coefficients as $\mathcal{T}^{(0)}$.
\item
We accomplish the circuit calculations by solving the set of the Eqs.~\eqref{eq:curr_4T_itg_set} and~\eqref{eq:EChP_4T}
using $\mathcal{T}^{(0)}$.
In this manner, which was in detail explained in Sect.~\ref{sssec:nonlin},
we obtain dependences $\mu^{(1)}(\VA,\VB)$, $\IA^{(1)}(\VA,\VB)$, and $\IB^{(1)}(\VA,\VB)$ as a first
approximation on chosen domains $\VA \in [0, \Vmax]$ and $\VB \in [0, \Vmax]$.
Later we refer to the (1)-superscripted results as the \iterprima level of approximation.
\item
Assuming that in further search we seek for refined $\mu^{(2)}(\VA,\VB)$, $\IA^{(2)}(\VA,\VB)$
and $\IB^{(2)}(\VA,\VB)$, we take the values $\mu^{(1)}(\VA,\VB)$ as new reference EChPs.
There are as many such references as is the number of the sampled points $(\VA,\VB)$ within the considered domain.
\item
The new reference (now non-equilibrium) potentials $\mu^{(1)}$ at a chosen particular $(\VA,\VB)$ point
are used for the next QM calculation of $\mathcal{T}_{\alpha\leftarrow\beta}(\eigE; \mu)$,
now denoted as $\mathcal{T}^{(1)}$.
\item
$\mathcal{T}^{(1)}$ is used in a new circuit calculation [again solving the Eqs.~\eqref{eq:curr_4T_itg_set}
and~\eqref{eq:EChP_4T}] thus yielding more accurate (\itersecunda) results $\mu^{(2)}$, $\IA^{(2)}$ and $\IB^{(2)}$,
now assumed to be valid at the chosen reference $(\VA,\VB)$ point only.
\end{enumerate}
Steps (4) and (5) are accomplished for a chosen subset of the grid points $(\VA,\VB)$ within the considered domain.
Next we can continue to improve the results at each chosen point $(\VA,\VB)$ performing \itertertia step, which takes
the $\mu^{(2)}$ potentials as a refined reference quite analogously to steps~(4) and~(5) of the above algorithm.
The number of such recursive iterations may be arbitrary.
To reach self-consistency of the currents and the EChPs,
only few such iterations will be needed as it will be seen in our numerical demonstrations
in Sect.~\ref{sssec:NonEqDemo}.

The QM calculation of the transmission coefficients requires the knowledge of the electrostatic potential profile
in the entire system including the graphene nanoribbon.
Within the independent-electron approximation we have to define a model spatial profile of the electrostatic potential
in the nanojunction.
As specified in Sect.~\ref{sec:model}, we employ the extended tight-binding hamiltonian
(see also Ref.~\cite{Konopka2015}) in which the spatial profile of the potential is expressed in terms of the diagonal
matrix elements of the hamiltonian (the on-site energies).
Assuming known values of the EChPs at the GNR's corners (we assume the same values as in the
corresponding electrodes), we employ the bilinear-interpolation formula, which defines values of the on-site energies
at any $(x, y)$ point of the GNR.
In more detail, the on-site energy of an atom $l$ positioned at point $(x, y)$ of the GNR is calculated as
$\epsilon_l = \epsilon + \Delta\epsilon_l$, where $\epsilon$ is the equilibrium on-site energy
and $\Delta\epsilon_l$ is the bias-induced modification computed using the bilinear interpolation
between the four corners.
$\Delta\epsilon_l$ could in addition include the effect of a gate voltage.
%
%
\section{\label{sec:applicate}Application to graphene nanoribbons}
%
%
We provide an illustrative set of results for the nanoribbon junction drawn in Fig.~\ref{fig:GNR}.
Two variants are considered:
(i)~The perfect GNR and (ii)~the GNR with the removed strip of the $60$ atoms marked in the figure by the dark narrow
strip, with the detail visualised in the inset of Fig.~\ref{fig:VB_IA_IB_Equilib_Lin}(b).
%
\subsection{\label{ssec:LinRespNeut}Coupled circuits in the linear-response regime at the neutrality point}
%
In this regime very low bias voltages and zero gate voltage are assumed.
We present results for both the intact GNR and the perturbed one with the cut-out.
Tab.~\ref{tab:Tmat}, which is placed within this section, lists also data computed for
a gate voltage scenario, which will be considered in Sect.~\ref{ssec:ResonRegime}.
\subsubsection{\label{sssec:intactGNR}The perfect GNR}
Using the microscopic approach described in Sect.~\ref{sec:model} we obtain numerical values of the linear conductance
for the perfect (intact) GNR
shown in Fig.~\ref{fig:GNR}.
The quantum subsystem of the whole scheme (Fig.~\ref{fig:scheme}) -- the GNR with its four
contacts -- provides a purely ballistic transport as we assume the low temperature limit.
The linear-response electronic transport properties of the junction under the equilibrium conditions
are then represented by a matrix $\mathcal{T}$ of the transmission coefficients,
or, alternatively, in terms of the corresponding conductance matrix $G = (2e^2/h)\,\mathcal{T}$.
The matrix is symmetric because of the absence of a magnetic flux.
Additional symmetries
($G_{12} = G_{34}$, $G_{13} = G_{24}$, $G_{14} = G_{23}$ and $G_{11} = \dots = G_{44}$)
follow from the $D_{2h}$ geometrical symmetry of the junction.
Off-diagonal matrix elements for the intact GNR are listed in the first
numerical row of Tab.~\ref{tab:Tmat} using the units of the conductance quantum $G_0 = 2 e^2/h$.
%
\begin{table}[t]
\caption{Off-diagonal elements $G_{\alpha\beta}$ of four linear-response conductance matrices in the units of $G_0$.
The first row of the numerical values describes the nanojunction based on the intact (perfect) GNR (Fig.~\ref{fig:GNR})
in the full equilibrium, i.e. without any bias or gate voltage applied.
The second row describes the case of the perturbed GNR, which is specified by Fig.~\ref{fig:GNR} as well, with a detail
shown by the inset to Fig.~\ref{fig:VB_IA_IB_Equilib_Lin}, again in the full equilibrium.
Rows 3 and 4 provide linear conductances for the same two GNRs with the gate voltage applied upon them
so that the transmission
is resonantly enhanced due to the ZZ edge states (Sect.~\ref{ssec:ResonRegime}).
The unlisted off-diagonal elements can be deduced from symmetries.}
\vskip12pt
\centering
\begin{tabular}{ld{1.5}d{1.5}d{1.5}d{1.5}}
\hline
& \multicolumn{1}{c}{$G_{12}$}
& \multicolumn{1}{c}{$G_{13}$}
& \multicolumn{1}{c}{$G_{14}$}
& \multicolumn{1}{c}{$G_{24}$}\\
\hline
Intact		& 0.1840	& 0.1246	& 0.1816	& 0.1246\\
Perturbed	& 0.1269	& 0.02575	& 0.1297	& 0.08584\\
Intact+$\Vg$	& 1.012		& 0.07974	& 0.1567	& 0.07974\\
Perturbed+$\Vg$	& 0.9180	& 0.07857	& 0.1859	& 0.1679\\
\hline
\end{tabular}
\label{tab:Tmat}
\end{table}
%
The ordering of the matrix elements is given by the indices $1, \dots, 4$ at the contacts of the GNR as defined
by Figs.~\ref{fig:GNR} and~\ref{fig:scheme}.
Not shown in the table are the diagonal elements.
They take values typically around $53$;
the particular intact GNR in the full equilibrium yields $\mathcal{T}_{\alpha\alpha} = 53.51$ for all four terminals
due to the $D_{2h}$ symmetry.
The large diagonal matrix elements can easily be understood for they mainly represent transmissions of the incoming
electrons back to the wires belonging to the same electrode.
In our model (Fig.~\ref{fig:GNR}) we assume $54$ (monoatomically thin) wires per electrode,
in other words $54$ conductance channels per electrode.
In terms of $\mathcal{T}$, any incoming electron close to the Fermi level is most likely to be transmitted to a wire
belonging to the same electrode what is the reason that the cumulative diagonal elements are all close to $54$.

In our first computational analysis of the entire scheme we use resistances
$\RA = 7.3\,\hbar/e^2 \approx 30.0\,\kO$ and
$\RB = 4.9\,\hbar/e^2 \approx 20.1\,\kO$.
%
\begin{figure}[t]
\centerline{\includegraphics[width=0.49\columnwidth]{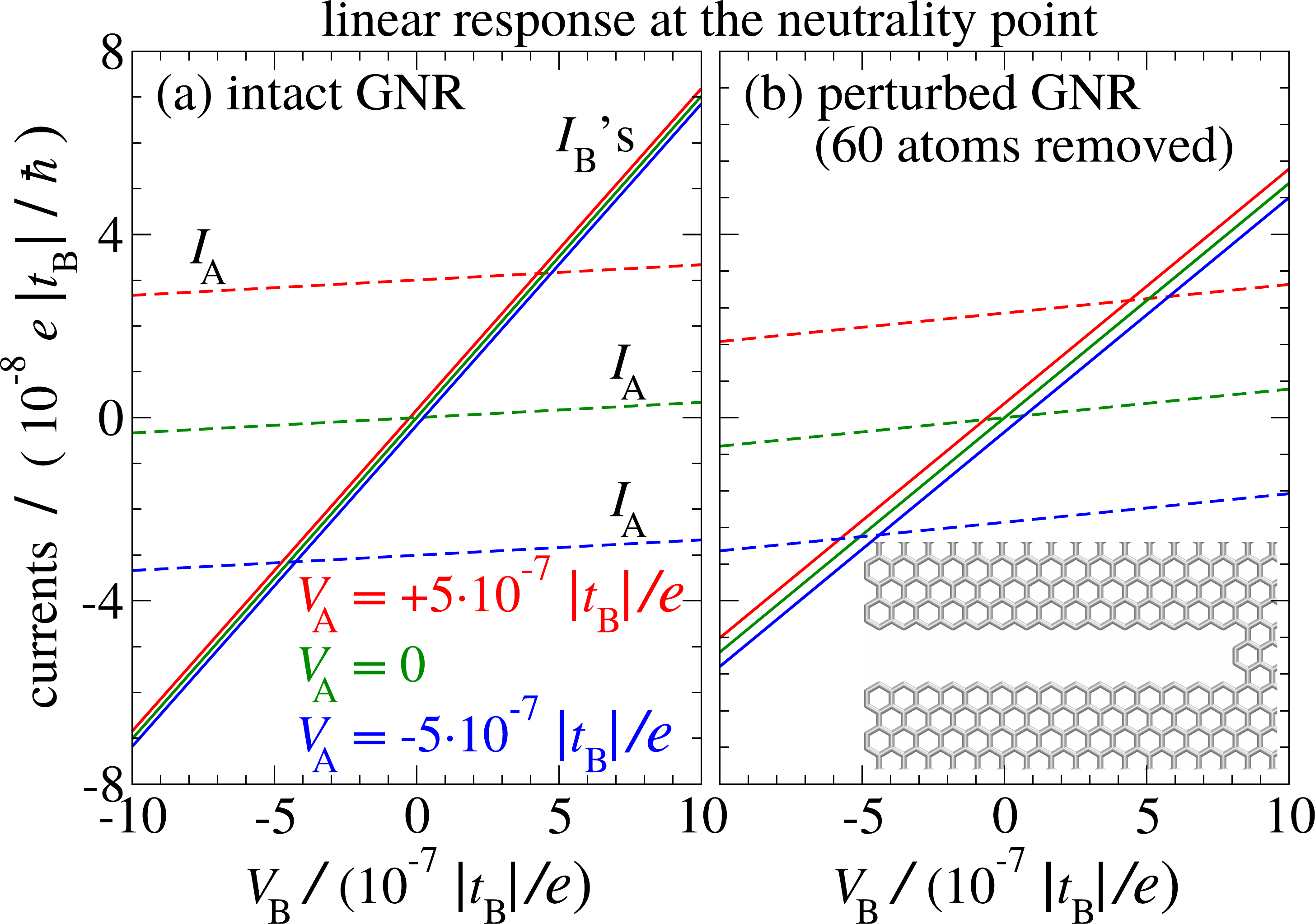}}
\caption{Illustration of the results in the linear regime for the two nanoribbon junctions
forming the devices according to the scheme in Fig.~\ref{fig:scheme}.
Conductances at the neutrality point (no gate voltage) are assumed.
(a)~Results for the perfect GNR shown in Fig.~\ref{fig:GNR}, with the conductance matrix specified by the
first numerical row of Tab.~\ref{tab:Tmat}.
(b)~Results for the perturbed GNR
with the conductance matrix elements given in the second numerical row of Tab.~\ref{tab:Tmat}.
The legends apply to both graphs.
The plots, according to the Eqs.~\eqref{eq:IA_IB_vs_VA_VB_lin}, show how the currents in the circuits A and B
vary with $\VB$ at the three fixed values of $\VA$.
The resistors used in the circuits have values
$\RA = 7.3\,\hbar/e^2 \approx 30.0\,\mathrm{k}\Omega$ and $\RB = 4.9\,\hbar/e^2 \approx 20.1\,\Omega$.
The natural voltage unit $|\tB|/e \approx 2.97\,\mathrm{V}$,
$e$ being the unit charge and $\tB$ the graphene nearest-neighbor hopping parameter.
In SI units the voltage $\VB$ scales up to $2.97\,\si{\micro\volt}$
while the currents axes scale up to $58\,\mathrm{pA}$.
The inset in graph~(b) shows the detail of the cut-out of the perturbed GNR that is
more thoroughly described by Fig.~\ref{fig:GNR} and in its caption.}
\label{fig:VB_IA_IB_Equilib_Lin}
\end{figure}
%
A typical or roughly an average value of the off-diagonal matrix elements for the equilibrium-level
conductance (the first row of Tab.~\ref{tab:Tmat}) is $\Goff \approx 0.16\,G_0  \approx  0.05\,e^2/\hbar$.
Our choice of the magnitudes of the resistances then corresponds to the regime with the classical conductances
$1/R$ of a similar order of magnitude as those of the quantum subsystem.
The results are conveniently visualised in terms of currents and are shown in Fig.~\ref{fig:VB_IA_IB_Equilib_Lin}(a).
The tiny magnitudes of the bias voltages and consequently of the currents (below $60\,\pA$) ensure the strictly linear
regime.
As can be seen in the figure, the current through circuit B is weakly influenced by the current in circuit A.
In other words, the two circuits influence each other to a minor although noticeable extent.
The seemingly simple, almost independent operation, of the two circuits is in fact a results of a cancelation
of the internal currents flowing inside the GNR between its terminals.
This can be easily understood using the formulae~\eqref{eq:IA_IB_vs_VA_VB_lin} for the currents together with
the expressions for B\"{u}ttiker's conductances $\alpha_{CC'}$~\cite{Buttiker1986}.
In terms of current $\IB$, the term responsible for the mutual interaction of the two circuits is proportional to
$-\alfaBA\VA$.
For the given highly symmetrical sample the conductance
$\alfaBA = (G_{31} G_{42} - G_{41} G_{32})/S =$ $(G_{31}^2 - G_{41}^2)/S \approx -0.00908\,e^2/\hbar$ 
is a very small quantity compared to $\alfaAA \approx 0.107\,e^2/\hbar$
as can be calculated from the matrix elements in the first row of Tab.~\ref{tab:Tmat}.
In other words, because $G_{31}$ and $G_{41}$ are of similar magnitudes,
they cause a mutual partial cancellation of the currents
flowing along the AC edges and the currents across the diagonal of the perfect GNR.
\subsubsection{\label{sssec:perturbedGNR}The perturbed GNR}
Electrical conductance properties of GNRs can be affected by different kinds of defects or designed features.
For example, the conductance of GNRs with the current flowing along their AC edges can be destroyed by the
AC edge disorder~\cite{White07}.
If a defect or a designed feature is such that especially the matrix elements $G_{31}$ and $G_{42}$
(or only one of them) are suppressed, then the cancellation effect discussed above would not work and,
consequently, currents in the two circuits would become influencing each other
in a more significant extent.
For the purpose of a quantitative illustration we consider a perturbed GNR
that was formed by the removal of the 60 atoms (15 quartets) from the perfect GNR according to the pattern
visualised in Figs.~\ref{fig:GNR} and~\ref{fig:VB_IA_IB_Equilib_Lin}(b) and described by their captions.
We note in passing that using the one-orbital-per-atom model we assume that all the edges of the structures
are passivated by hydrogens~\cite{White07}.
The linear conductance of the perturbed GNR at the neutrality point is described by the matrix elements,
most important of which are shown in the second numerical row of Tab.~\ref{tab:Tmat}.
The matrix element $G_{13}$ is most significantly suppressed in the relative terms.
The electric currents resulting from this conductance matrix are plotted in pane (b) of
Fig.~\ref{fig:VB_IA_IB_Equilib_Lin} at otherwise the same conditions as were assumed for pane~(a).
As can be inspected, pane (b) demonstrates the regime in which the two classical circuits interact through the quantum
device more significantly [compared to pane (a)]:
currents in circuit B depend on voltage in circuit A more strongly and also
currents $\IA$ depend more strongly on voltage $\VB$.

The suppression of the $G_{13}$ matrix element might look intuitively obvious, given the location of the vacant strip
in the GNR.
Quantitative calculations however show that such a picture is not obvious at all;
a removal of, for instance, one quartet of atoms fewer from the perfect
GNR (thus leaving the vacant strip correspondingly shorter)
would yield a structure with $G_{13}$ about 5 times higher than that for the perfect GNR.
Continued removal of more and more 4-atom groups from the GNR (making the vacant strip step by step longer)
would yield a sharply oscillating pattern of the conductance characteristics as we have verified.
Although such an analysis is not a subject of the present work,
we note that similar behaviour is well-know for armchair GNRs which, depending on their width, exhibit three
different conductance patterns~\cite{Nakada1996,Louie_PRL06}.
We chose the particular perturbed structure with its conductance matrix in order to demonstrate the emergence
of the more significant correlations between the currents $\IA$ and $\IB$ which can occur if the above discussed
mutual cancellation effect of the internal currents (Sect.~\ref{sssec:intactGNR}) is reduced.
A more detailed description and quantitative results for the GNRs with this type of the structural perturbation
can be found in the SM~\cite{SM}.
\subsubsection{\label{sssec:balance}Balancing the Fermi levels}
As it can be directly observed from the results of Sect.~\ref{sssec:lin_nospin},
the effects of the resistors are significant and most interesting in the regime with the resistances of the magnitudes
$R \approx 1/\Goff$ (semi-quantitatively).
%
\begin{figure}[!t]
\centerline{\includegraphics[width=0.35\columnwidth]{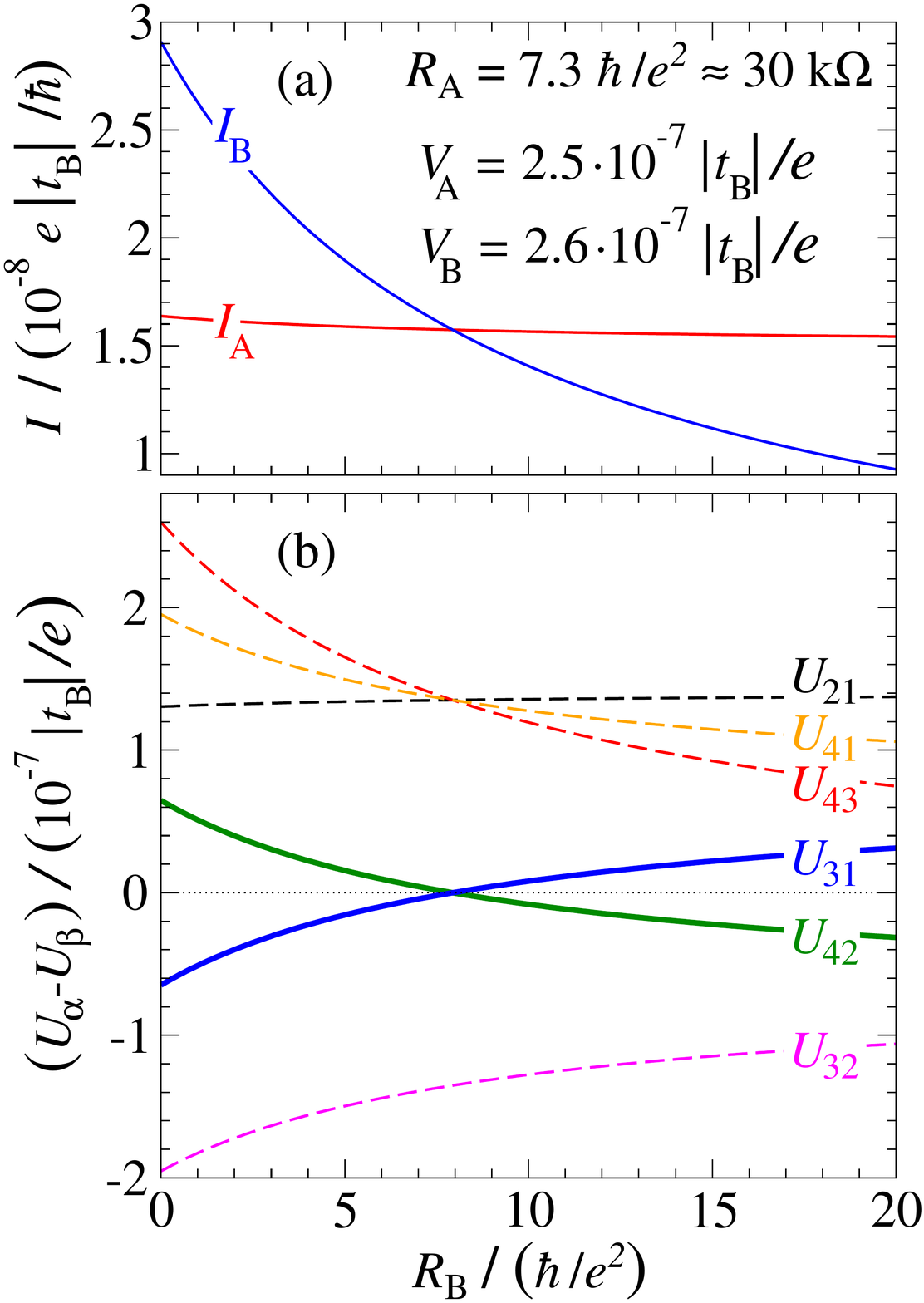}}
\caption{Illustration of the leads' quasi-Fermi level balancing using a tunable resistor
(the $U_{42}$ and $U_{31}$ curves).
The perfect GNR and the linear-response neutrality-point conductance are assumed.
The electric currents~(a) and the electrostatic potentials differences
$U_{\alpha\beta} \equiv U_\alpha - U_\beta$~(b) resulting from the
formulae~\eqref{eq:IA_IB_vs_VA_VB_lin}, \eqref{eq:U42_solo} and from similar formulae
for the other $U_{\alpha\beta}$'s~\cite{SM} are plotted as functions of the resistance $\RB$.
The resistance $\RA$ and the bias voltages $\VA$ and $\VB$ remain fixed at their values typed within the upper pane.
The relevant conductance matrix elements are provided in the first row of Tab.~\ref{tab:Tmat}.
The figure illustrates mainly the regime of (semi-quantitatively) $R \approx 1/G^\mathrm{off}$, where $G^\mathrm{off}$
is a typical value of a non-diagonal element of the conductance matrix.
The thick green curve crosses the zero level thus indicating that at that point $U_2 = U_4$ or that the EChPs
at leads 2 and 4 match (see also Fig.~\ref{fig:scheme}).
Due to the high symmetry of the studied GNR the potentials $U_1$ and $U_3$ match at the same value of $\RB$.
For conversions to SI units the value of $|\tB| \approx 2.97\,\eV$ should be used as specified in the text.}
\label{fig:U42etc}
\end{figure}
%
Employing a tunable resistor (or both of them) in this regime, one can balance the quasi-Fermi levels~\eqref{eq:mu}
of a certain pair of the leads, provided that the remaining physical parameters permit it.
One particular scenario is described by the Eq.~\eqref{eq:U42_solo}.
An illustration of the effect for the intact GNR is shown in Fig.~\ref{fig:U42etc} in which the currents and the
potentials differences are plotted as given by the Eqs.~\eqref{eq:IA_IB_vs_VA_VB_lin},
\eqref{eq:U42_solo} and by the formulae for the other $U_{\alpha\beta}$'s~\cite{SM}.
We keep $\RA = 7.3 \, \hbar/e^2 \approx 30.0\,\kO$.
The bias voltages on the sources use the values
$\VA = 2.5 \cdot \! 10^{-7}\,|\tB|/e \approx 0.74 \cdot \! 10^{-6}\,\si{\micro\volt}$
and
$\VB = 2.6 \cdot \! 10^{-7}\,|\tB|/e \approx 0.77 \cdot \! 10^{-6}\,\si{\micro\volt}$.
The electrostatic potentials of leads 2 and 4 match at the value
$\RB^{\mathrm{match}} = 7.936\,\hbar/e^2 \approx 32.6\,\mathrm{k}\Omega$;
at this value the solid dark green curve crosses zero thus indicating the equality $U_2 = U_4$.
Because of the high symmetry of the studied sample (the intact GNR), the potentials $U_1 \equiv 0$ and $U_3$
(the solid blue plot) are balanced at the same value of $\RB$.
%
\subsection{\label{sssec:NonEqDemo}Non-linear response at non-equilibrium Fermi levels}
%
If the magnitude of one or both of the bias voltages $\VA$ and $\VB$ is increased above certain threshold
%
\begin{figure}[!t]
\centerline{\includegraphics[width=60mm]{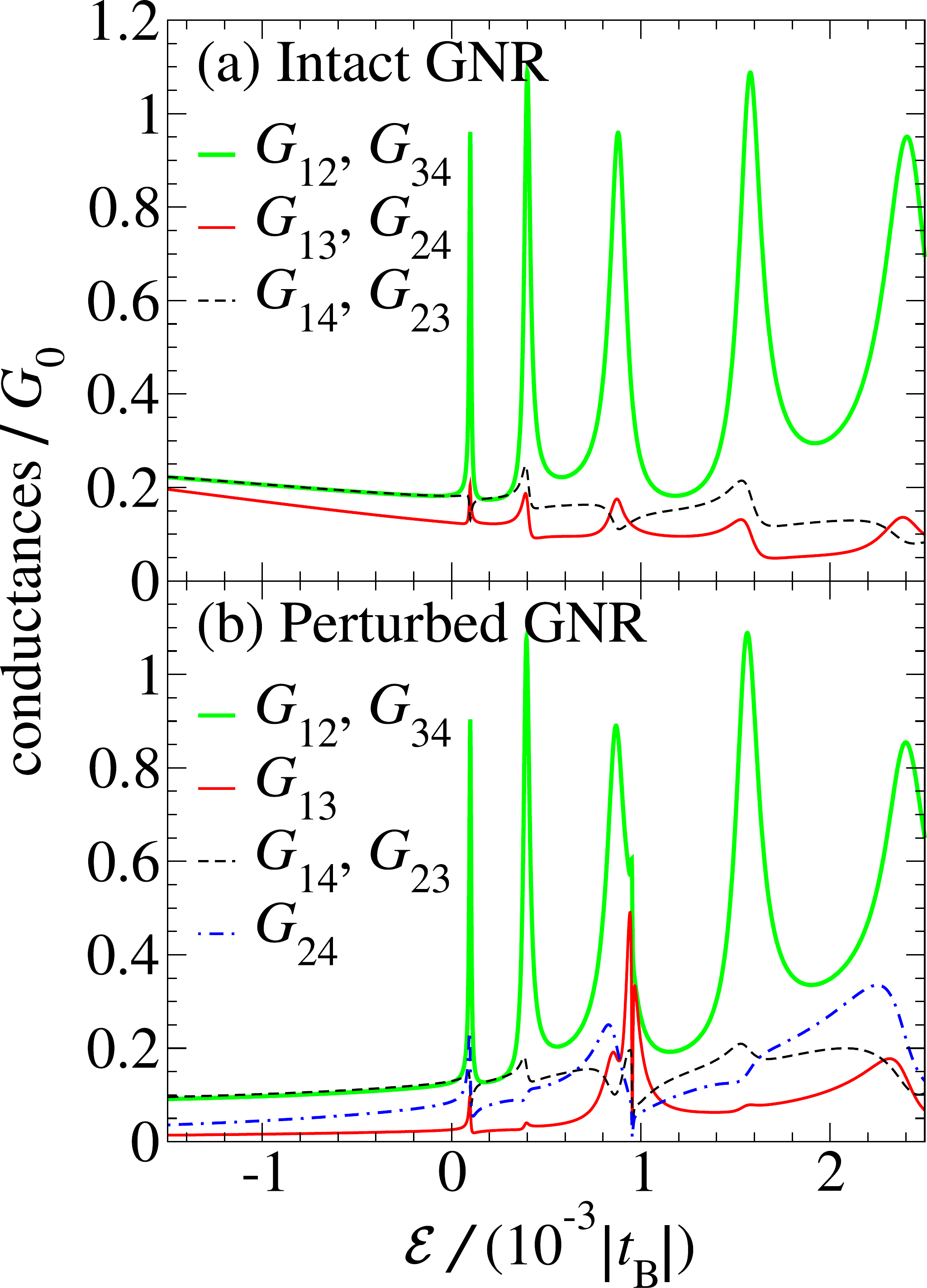}}
\caption{The elements of the conductance matrix $G_{\alpha\beta}$ expressed in the units of the conductance quantum as
functions of the energy for the
(a)~intact (perfect) GNR shown in Fig.~\ref{fig:GNR},
(b)~perturbed GNR shown again in Fig.~\ref{fig:GNR} and in the inset of Fig.~\ref{fig:VB_IA_IB_Equilib_Lin}.
The spectra have been calculated at the full-equilibrium conditions.
Their values for the intact structure at $\eigE = 0$ are given by the first row in Tab.~\ref{tab:Tmat}.}
\label{fig:Gcond_vs_E}
\end{figure}
%
(around $10^{-4}\,|\tB|/e$ in our model), 
the EChPs $\mu_2$, $\mu_3$, and $\mu_4$ may take values such that
the electrons most relevant for the transport acquire energies from the range of the ZZ edge
modes.\footnote{Particular values of the electrodes' EChPs at which the ZZ-edge modes are probed
depend on an applied gate voltage that can effectively shift the energy levels of the central device.
Apart from Sect.~\ref{ssec:ResonRegime}, zero gate voltage (no gate field) is assumed.}
%
Conductance matrix $G_{\alpha\beta}$ is significantly energy-dependent in this energy window.
Fig.~\ref{fig:Gcond_vs_E} shows the spectra of the matrix elements computed under the full-equilibrium conditions
(in particular at zero gate field).
The solid green lines represent the matrix elements strongly affected by the presence of the zigzag edges of the GNR.
%
\begin{figure}[!t]
\centerline{\includegraphics[width=100mm]{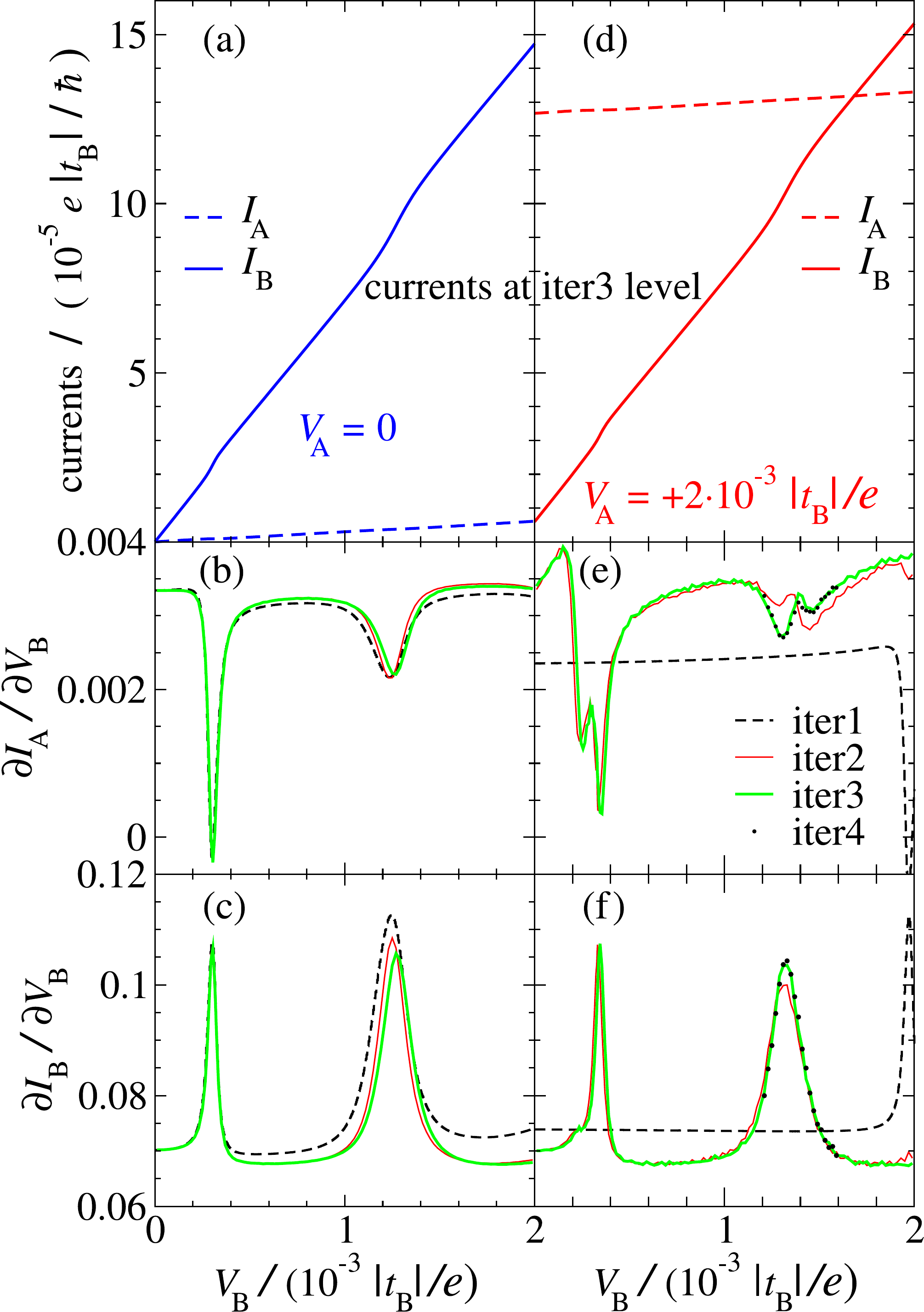}}
\caption{Results for the non-linear response regime
with the iterative account of the non-equilibrium Fermi levels
(Sects.~\ref{sssec:NonEqTheory} and~\ref{sssec:NonEqDemo}).
The intact GNR (Fig.~\ref{fig:GNR}) and zero gate field are assumed.
The resistors in the circuits take the same values as were used for Fig.~\ref{fig:VB_IA_IB_Equilib_Lin}.
The left-hand side panes (a, b, c) assume zero $\VA$ bias voltage
while the right-hand side ones (d, e, f) show results obtained at $\VA = 0.002\,|\tB|/e$.
(a, d):~I/V plots similar to those in Fig.~\ref{fig:VB_IA_IB_Equilib_Lin} but now up to higher (an non-negative only)
bias voltages $\VA$ and $\VB$.
The differential conductances (b, c, e, f) were computed at the three or four subsequent iterations and are shown
in the units of $e^2/\hbar$.
The subset of the results provided at the \iterquarta level
[discrete black dots in panes (e) and (f)] demonstrates that already the $3^\mathrm{rd}$ iteration was sufficient.}
\label{fig:NL_NonEq_regime}
\end{figure}
%
The sharp oscillations due to the ZZ-edge modes contribute to the currents and make them non-linear
with the increasing magnitudes of the the bias voltages $\VA$ and $\VB$.
The oscillations start at energies as low as about $8 \cdot \! 10^{-5}\,|\tB|$ above the equilibrium Fermi level
meaning that the transport becomes non-linear even for such relatively low bias voltages.
The related theory and computational procedures used to analyse the composite quantum-classical system
(Fig.~\ref{fig:scheme}) in the non-linear regime assuming a \emph{single set of spectra} $G_{\alpha\beta}(u)$
were exposed in Sect.~\ref{sssec:nonlin}.
That level of approximation (denoted as \iterprima in Sect.~\ref{sssec:NonEqTheory}) does not in general provide
sufficiently accurate results for the model under study.
It fails especially for more sensitive quantities like differential conductances.
Still, it often provides acceptable estimates for currents $\IA$ and $\IB$ and for the EChPs.
A set of \iterprima results, in comparison to the linear response regime, can be found in the SM~\cite{SM}.

In Fig.~\ref{fig:NL_NonEq_regime}(a,d) we present non-linear I/V characteristics,
which, in addition, include the effects of the non-equilibrium EChPs
using the iterative scheme described in Sect.~\ref{sssec:NonEqTheory}.
The intact (perfect) GNR at zero gate field is assumed
and the $\IA(\VB)$ and $\IB(\VB)$ curves are obtained for the two fixed values of $\VA$.
Because the $I_\mathrm{A,B}$ vs. $\VB$ plots themselves do not display their non-linear features in a pronounced way,
the differential conductances $\partial\IA/\partial\VB$ and $\partial\IB/\partial\VB$
are more convenient for this purpose.
We visualise them in graphs (b), (c), (e) and (f) of Fig.~\ref{fig:NL_NonEq_regime} using the units $e^2/\hbar$
(i.e. not $G_0$).
Main results in these panes are outlined using the green thick solid plots (\itertertia).
The additional curves demonstrate how the recursive calculation converges toward self-consistency of the 
currents and the EChPs:
three iterations of the scheme described in Sect.~\ref{sssec:NonEqDemo} were sufficient to obtain the converged results
(the green thick solid plots).
This is seen from panes (e) and (f) in which we provide a subset of results at the \iterquarta level
(the discrete black points).
%
They perfectly agree with the \itertertia plots.\footnote{\iterprima differential conductances were obtained using
the numerical solution and analytical formulae based on the description in Sect.~\ref{sssec:nonlin}, with additional
details provided in~\cite{SM}.
The derivatives for the higher iterations were obtained by direct numerical differentiation of the currents.
The sampling density for the higher iterations was $2.0 \cdot 10^{-7}\,|\tB|/e$.
Since the solutions were accomplished on the finite-density numerical grid, certain finite-size errors are necessary;
we do not employ any smoothening procedure and show the results as they come from the calculations.
We have verified that our numerical procedures are stable and a doubling of the grid density would not present
any practical difficulty and would yield smoother results.}
%
In the SM~\cite{SM} we provide a direct comparison of several subsequent iterations also for the currents $\IA$
and $\IB$.
As a complement we outline there also the EChPs $\mu_{\alpha}(\VB)$.
%
\subsection{\label{ssec:ResonRegime}Resonantly increased conductance in the linear regime}
%
In Sect.~\ref{sssec:perturbedGNR} the impact of the particular perturbation of the originally perfect GNR was examined
in the low-voltage regime around the neutrality point of the studied model.
[Cf. pane (b) vs.~(a) of Fig.~\ref{fig:VB_IA_IB_Equilib_Lin}.]
The cut-out of the 60 atoms affected especially the $G_{13}$ matrix element, magnitude of which was
suppressed to about $0.207$ of the original value (Tab.~\ref{tab:Tmat}).
The \emph{effectively} almost independent conductance channels [Fig.~\ref{fig:VB_IA_IB_Equilib_Lin}(a)]
along the two ZZ edges then turned to the more significantly coupled ones [Fig.~\ref{fig:VB_IA_IB_Equilib_Lin}(b)].

\begin{figure}[t]
\centerline{\includegraphics[width=0.49\columnwidth]{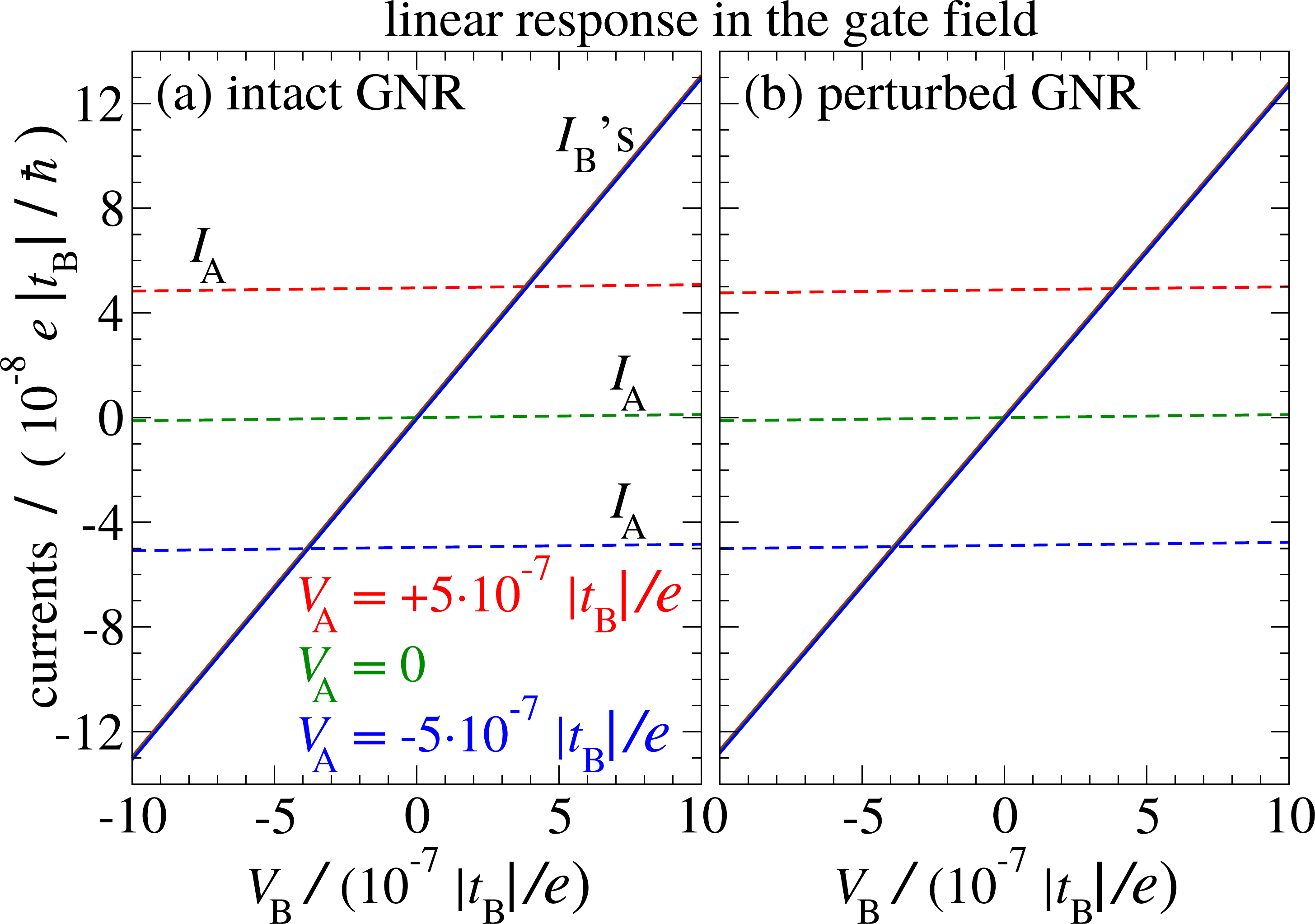}}
\caption{Linear I/V plots calculated for the resonantly increased conductance due to the applied gate voltage such
that it shifts the atomic orbital energies of the GNR by $-0.0016\,|\tB|$ and the conductance matrix is given by
the third and fourth numerical rows of Tab.~\ref{tab:Tmat}.
All other conditions (including the number of displayed plots)
are identical to those use for Fig.~\ref{fig:VB_IA_IB_Equilib_Lin}.
The $\IB$ plots are now very close each other so that they are indistinguishable on the graphs.}
\label{fig:VB_IA_IB_Reson}
\end{figure}

One way how to achieve a regime of the relatively independent currents along the ZZ edges regardless of possible
disruption or disorder of the AC edges, is to increase the conductance along the ZZ edges.
This can be achieved using resonant transport conditions such that the ZZ edge modes are employed.
Such a tuning to the resonance can be obtained using a proper gate field applied to the GNR while still working
at tiny bias voltages, i.e. in the linear regime.
In this way, upon applying a gate voltage such that it shifts the orbital energies of the GNR by $-0.0016\,|\tB|$,
we obtain conductance matrices\footnote{The effect of such a gate voltage for our models is largely just to shift the
spectra displayed in Fig.~\ref{fig:Gcond_vs_E};
either we perform an explicit QM calculation with an applied gate voltage or we just shift the spectra,
we obtain practically the same matrices, with a relative difference not larger than $10^{-4}$.}
with their elements shown in the third and fourth numerical rows of Tab.~\ref{tab:Tmat},
which should be contrasted to the matrices obtained at the neutrality point
(the first and second rows).
The $G_{12}$ column of Tab.~\ref{tab:Tmat} clearly demonstrates that the junctions under the gate voltage
provide greatly enhanced conductance along the ZZ edges.
As for the other matrix elements in Tab.~\ref{tab:Tmat}, variations of their particular values may not be obvious.
We have briefly commented on this point at the end of Sect.~\ref{sssec:perturbedGNR},
with the link to the SM~\cite{SM}.
A graphical representation of the results in terms of the currents, now for the resonantly increased conductances
$G_{12} = G_{34}$, may again be convenient:
we prepare the two-pane Fig.~\ref{fig:VB_IA_IB_Reson} in the same style as Fig.~\ref{fig:VB_IA_IB_Equilib_Lin}.
For easier comparisons we employ the same values of the bias voltages $\VA$, $\VB$ and of the resistances $\RA$, $\RB$
as were assumed for Fig.~\ref{fig:VB_IA_IB_Equilib_Lin}.
%
%
The vertical axes are now scaled up to higher values resulting predominantly from the resonantly increased matrix
elements $G_{12} = G_{34}$.
The increase of the currents in comparison to Fig.~\ref{fig:VB_IA_IB_Equilib_Lin}
is not such as might be expected from the significantly larger matrix elements $G_{12}$.
We recall that the classical resistances $\RA$ and $\RB$ in the circuits are assumed to be the same (and significant)
as they were for Fig.~\ref{fig:VB_IA_IB_Equilib_Lin} and they consequently limit the currents.
Nevertheless, the relatively large $G_{12}$ numerical values for the junction in the gate potential imply the
dominance of the transport along the ZZ edges.
The operation of the two circuits is then expected to be mutually almost independent as opposed to the case shown
in Fig.~\ref{fig:VB_IA_IB_Equilib_Lin}(b).
Graphical representations of the results in both panels of Fig.~\ref{fig:VB_IA_IB_Reson} confirm the expectation:
the currents in the circuit B practically do not depend on the voltage applied in the circuit A, at least for
the range of the voltages assumed for the graph.
Analogously, the dependence of $\IA$'s on $\VB$ is similarly weak for the gated junctions.
In this way the ZZ edges can serve as two almost uncoupled conductors for DC low-voltage transport.
In the SM~\cite{SM} we provide more comprehensive results and comparisons including two additional perturbed
structures.
%
\section{\label{sec:concl}Conclusions}
%
%
We considered a coupled quantum-classical device consisting of an atomistic model of a nanoribbon (NR)
and of two classical branches of circuits with resistances and bias voltage sources.
The four electrodes of the circuits were assumed to be contacted at the corners of the NR so that the
edge-induced modes could be used for the electrical transport.
The two bias voltages, not electrochemical potentials (EChPs), has been assumed as the control variables.
In order to analyse the device operation, we have developed a model, which allowed us to calculate the two electric
currents and three non-trivially dependent EChPs in the leads as functions of the two externally
controlled bias voltages, with the NR's conductance matrix (in general energy dependent) and the classical resistances
as given parameters of the model.
The model is usable for a general four-terminal ballistic device and presents a generalisation of B\"{u}ttiker's
model~\cite{Buttiker1986}, which assumes zero resistances in the circuits and is limited to the linear-response regime.
As a particular material for our computational analysis we assumed graphene.
We calculated the conductance matrix using a microscopic theory based on the Green's function matrix formalism.
Our present implementation of the formalism allowed us to consider graphene nanoribbons (GNRs) up to a size of
about $30\,000$ atoms with computational resources of a desktop personal computer.

In the low-voltage linear regime we obtained transparent analytical formulae for the currents
$I_{\mathrm{A},\mathrm{B}}$ and electrostatic potentials $U_\alpha$ in the system as functions of the applied
bias voltages $V_{\mathrm{A},\mathrm{B}}$;
hence also formulae for the EChPs $\mu_\alpha = -e U_\alpha$ of the leads.
The formulae for the currents generalise those found by B\"{u}ttiker~\cite{Buttiker1986}.
Our formulae show how the interplay of the quantum and classical elements of the entire scheme affects the resulting
electric currents.
Similarly, our formulae for the electrochemical (or electrostatic) potentials at the device's terminals display the
non-trivial dependences of the potentials on the classical resistances and on the quantum conductance matrix elements.

For the voltages in the non-linear regime we developed an efficient numerical algorithm to compute
the $I/V$ and $\mu/V$ curves.
The non-equilibrium effects have been included using a recursive algorithm in which self-consistency
between the EChPs and the currents is reached within few iterations.

Our analysis has shown under what conditions the two edges of a nanoribbon behave as two relatively independent wires
transporting stationary electric currents.
On the other hand, the intriguing regime in which the two classical circuits are significantly coupled through the
quantum ballistic device has also been studied:
an increase of the voltage (and of the current) in one of the circuits induces a substantial change in the current
in the other circuit.

Quantitative results have been demonstrated for a GNR with zigzag (ZZ) edge induced high density of states close to the
Fermi level.
The system was found to behave similarly to two independent conductors especially for resonantly enhanced conductance
along the ZZ edges.
For other regimes the mutual interaction of the two associated circuits was found noticeable.
Results have been shown for the regime $1/\RAB \lesssim \Goff$
($\RAB$ - the resistances in the classical subsystem,
$\Goff$ - a typical value of an off-diagonal conductance element of the quantum subsystem -- the nanoribbon).
Although we have considered relatively small GNRs with numbers of atoms of the order of $10^3$, our model can be
applied to any junction supporting ballistic transport of electrons for which its conductance matrix is know.
For example, larger-scale devices such as graphene ribbons used in Ref.~\cite{Allen_2016} could be described as
well provided that the conductance matrix could be computed or obtained from experimental measurements.
%
%
\section{Acknowledgements}
%
%
This work was supported in parts by
the Slovak Research and Development Agency under the contract No.~APVV-0108-11
and by
the Slovak Grant Agency for Science (VEGA) through grant No.~1/0372/13.
Useful discussions with Peter Bokes and Richard Hlubina are acknowledged.
The GNR structure was visualised using VMD~\cite{VMD}.

Contributions of the authors:
P.~D. did most of the analysis of the semi-infinite chains eigenstates presented in the SM~\cite{SM},
partially contributed to derive the compact forms~\eqref{eq:IA_IB_vs_VA_VB_lin} of the formulae for the
currents and derived the relations (38) and (39) in the SM.
All other contributions have been done by M.~K.


\begin{thebibliography}{99}

\bibitem{Nakada1996}
K. Nakada, M. Fujita, G. Dresselhaus, M.S. Dresselhaus,
\textit{Edge state in graphene ribbons: Nanometer size effect and edge shape dependence},
Phys. Rev. B \textbf{54}, 17954 (1996).

\bibitem{Yazyev2008}
O.V. Yazyev, M.I. Katsnelson,
\textit{Magnetic Correlations at Graphene Edges: Basis for Novel Spintronics Devices},
Phys. Rev. Lett. \textbf{100}, 047209 (2008).

\bibitem{Niu2009}
W. Yao, S.A. Yang, Q. Niu,
\textit{Edge States in Graphene: From Gapped Flat-Band to Gapless Chiral Modes},
Phys. Rev. Lett. \textbf{102}, 096801 (2009).

\bibitem{Allen_2016}
M.T. Allen, O. Shtanko, I.C. Fulga, A.R. Akhmerov, K. Watanabe, T. Taniguchi, P. Jarillo-Herrero, L.S. Levitov,
A. Yacoby,
\textit{Spatially resolved edge currents and guided-wave electronic states in graphene},
Nat. Phys. \textbf{12}, 128 (2016).

\bibitem{Dai2008}
X. Wang, Y. Ouyang, X. Li, H. Wang, J. Guo, H. Dai,
\textit{Room-Temperature All-Semiconducting Sub-10-nm Graphene Nanoribbon Field-Effect Transistors},
Phys. Rev. Lett. \textbf{100}, 206803 (2008).

\bibitem{Konopka2015}
M. Kon\^{o}pka,
\textit{Conductance of graphene flakes contacted at their corners},
J. Phys.: Condens. Matter. \textbf{27}, 435005 (2015).

\bibitem{Buttiker1986}
M. B\"{u}ttiker,
\textit{Four-Terminal Phase-Coherent Conductance},
Phys. Rev. Lett. \textbf{57}, 1761 (1986).

\bibitem{Wang1992}
Jian Wang, Yong Wang, Hong Guo,
\textit{Ballistic-electron transport through a coupled-quantum-wire system},
Phys. Rev. B \textbf{46}, 2420 (1992).

\bibitem{Hirayama1992}
Y. Hirayama, A. D. Wieck, T. Bever, K. von Klitzing, K. Ploog,
\textit{Parallel in-plane-gated wires coupled by a ballistic window},
Phys. Rev. B \textbf{46}, 4035 (1992).

\bibitem{Joachim_2002_PRB}
S. Ami, C. Joachim,
\textit{Intramolecular circuits connected to N electrodes using a scattering matrix approach},
Phys. Rev. B \textbf{65}, 155419 (2002).

\bibitem{Joachim_2002_Nanotech}
S. Ami, M. Hliwa, C. Joachim,
\textit{Balancing a four-branch single-molecule nanoscale Wheatstone bridge},
Nanotechnology \textbf{14}, 283 (2003).

\bibitem{Lesovik_1993}
G.B. Lesovik, C. Presilla,
\textit{Nonlinear voltages in multiple-lead coherent conductors},
Phys. Rev. B \textbf{47}, 2398 (1993).

\bibitem{SM}
Martin Kon\^{o}pka, Peter Die\v{s}ka,
the Supplementary Material to the present work.

\bibitem{Ruffieux2016}
S. Wang, L. Talirz, C.A. Pignedoli, X. Feng, K. M\"{u}llen, R. Fasel, P. Ruffieux,
\textit{Giant edge state splitting at atomically precise graphene zigzag edges},
Nat. Commun. \textbf{7}, 11507 (2016).

\bibitem{Reich2002}
S. Reich, J. Maultzsch, C. Thomsen, P. Ordej\'{o}n,
\textit{Tight-binding description of graphene},
Phys. Rev. B \textbf{66}, 035412 (2002).

\bibitem{Yazyev2010}
O.V. Yazyev,
\textit{Emergence of magnetism in graphene materials and nanostructures},
Rep. Prog. Phys. \textbf{73}, 056501 (2010).

\bibitem{our_PRB_2014}
M. Kon\^{o}pka, P. Bokes,
\textit{Wave-packet representation of leads for efficient simulations of time-dependent electronic transport},
Phys. Rev. B \textbf{89}, 125424 (2014).

\bibitem{Ryndyk}
D.A. Ryndyk, R. Guti\'{e}rrez, B. Song, G. Cuniberti,
\textit{Green Function Techniques in the Treatment of Quantum Transport at the Molecular Scale},
in \textit{Energy Transfer Dynamics in Biomaterial Systems, edited by I. Burghardt, V. May, D.A. Micha, E. Bittner},
Springer Series in Chemical Physics Vol. 93
(Springer-Verlag, Berlin, Heidelberg, 2009), p. 213.

\bibitem{Christen1996}
T. Christen, M. B\"{u}ttiker,
\textit{Gauge-invariant nonlinear electric transport in mesoscopic conductors},
Europhys. Lett. \textbf{35}, 523 (1996).

\bibitem{HaugJauho}
H. Haug, A.-P. Jauho,
\textit{Quantum Kinetics in Transport and Optics of Semiconductors},
Springer Series in Solid-State Physics Vol. 123
(Springer-Verlag, Berlin, Heidelberg, 2008).

\bibitem{White07}
D.A. Areshkin, D. Gunlycke, C.T. White,
\textit{Ballistic Transport in Graphene Nanostrips in the Presence of Disorder: Importance of Edge Effects},
Nano Lett. \textbf{7}, 204 (2007).

\bibitem{Louie_PRL06}
Y.-W. Son, M.L. Cohen, S.G. Louie,
\textit{Energy Gaps in Graphene Nanoribbons},
Phys. Rev. Lett. \textbf{97}, 216803 (2006).

\bibitem{VMD}
W. Humphrey, A. Dalke, K. Schulten,
\textit{VMD: Visual Molecular Dynamics},
J. Molec. Graphics \textbf{14}, 33 (1996).

\end{thebibliography}
\end{document}